\documentclass[12pt]{article}
\usepackage{graphicx}
\usepackage{cite}
\usepackage{color}
\usepackage{amssymb}
\newcommand{\bm}[1]{\mbox{\boldmath $#1$}}
\newcommand{\fnd}[2]{\frac{\textstyle #1}{\textstyle #2}}
\newcommand{\abs}[1]{\left| #1\right|}
\newcommand{\braket}[3]{\mbox{$\left\langle #1\left|
#2\right| #3\right\rangle$}}
\newcommand{\fndrs}[4]{\fnd{\raisebox{#1}{$#2$}}{\raisebox{#3}{$#4$}}}
\newcommand{\x}[1]{{\textstyle #1}}
\newcommand{\xrm}[1]{{\textstyle \mbox{\rm #1}}}

\newcommand{\lijntje}[3]{(\begin{picture}(20,0)(4,0)
\end{picture})}
\definecolor{pink}{rgb}{0.5,0.0,1.0}
\begin{document}
\title{Production of hadron pairs
in $e^{+}e^{-}$ annihilation\\
near the $K^{+}K^{-}$, $D\bar{D}$, $B\bar{B}$
and $\Lambda^{+}_{c}\Lambda^{-}_{c}$ thresholds}
\author{
Eef~van~Beveren$^{\; 1}$ and George~Rupp$^{\; 2}$\\ [10pt]
$^{1}${\small\it Centro de F\'{\i}sica Computacional,
Departamento de F\'{\i}sica,}\\
{\small\it Universidade de Coimbra, P-3004-516 Coimbra, Portugal}\\
{\small\it eef@teor.fis.uc.pt}\\ [10pt]
$^{2}${\small\it Centro de F\'{\i}sica das Interac\c{c}\~{o}es Fundamentais,}\\
{\small\it Instituto Superior T\'{e}cnico, Universidade T\'{e}cnica de
Lisboa,}\\
{\small\it Edif\'{\i}cio Ci\^{e}ncia, P-1049-001 Lisboa, Portugal}\\
{\small\it george@ist.utl.pt}\\ [.3cm]
{\small PACS number(s): 14.40.Gx, 14.40.Cs, 13.25.Gv, 11.80.Gw}
}

\maketitle
\begin{abstract}
We study the structures in cross sections and $R$ ratios concerning
$e^{+}e^{-}$ annihilation reactions, for open-strangeness, open-charm
and open-beauty hadron-pair production, from the perspective of a
recently developed production formalism. Special attention is paid to
non-resonant contributions. Known, tentative, and yet unknown resonances
are spotted, with estimates for their masses and widths.
\end{abstract}

\section{Introduction}

Several of the new enhancements in production cross sections share one common
property, namely, they occur at --- or just above --- an important threshold.
The most recent example, the $J/\psi\,\phi$ enhancement
observed and baptized $Y(4140)$ by the CDF Collaboration \cite{PRL102p242002},
appears right above the $J/\psi\,\phi$ threshold.
The enhancement in the $e^{+}e^{-}\to\Lambda_{c}^{+}\Lambda_{c}^{-}$ cross
section reported by the Belle Collaboration \cite{PRL101p172001}
occurs right above the $\Lambda_{c}^{+}\Lambda_{c}^{-}$ threshold.
The $Y(4260)$ enhancement in the $e^{+}e^{-}\to J/\psi\pi^{+}\pi^{-}$ cross
section, observed by the BaBar Collaboration \cite{PRL95p142001},
is right on top of the $D_{s}^{\ast}D_{s}^{\ast}$ threshold.
The $X(3872)$ enhancement \cite{PRL91p262001} in
$B^{\pm}\to J/\psi K^{\pm}\pi^{+}\pi^{-}$ decay
lies just above the $DD^{\ast}$ threshold.
The new enhancement found by the BES Collaboration at 3.763 GeV
\cite{ARXIV08070494} in the cross section of hadronic
electron-positron annihilation reactions shows up just above the $D\bar{D}$
threshold. Therefore, we expect a common explanation for these phenomena.

In the present work, we shall apply a recently developed formalism
for the study of hadronic electron-positron annihilation reactions
to determine common properties of the signal for open-strangeness, open-charm,
and open-beauty hadron-pair production near threshold.
The formalism is outlined in Sec.~\ref{formalities} and applied in
Sec.~\ref{production}. Conclusions are presented in Sec.~\ref{finalities}.

\section{Hadron-pair production in \bm{e^{-}e^{+}} annihilation}
\label{formalities}

In Ref.~\cite{AP323p1215}, we derived a relation between
the production amplitude $a_{\ell}(\alpha\to i)$s
for $e^{+}e^{-}$ to two-meson annihilation reactions
and the matrix elements of the meson-meson scattering amplitude $T$,
resulting for the $\ell$-th partial wave in an expression of the form
$T_{\ell}(\nu\to i)$
\begin{equation}
a_{\ell}(\alpha\to i)\propto
g_{\alpha i}\, j_{\ell}\left( p_{i}r_{0}\right)
+\frac{i}{2}\sum_{\nu}\,
h^{(1)}_{\ell}\left( p_{\nu}r_{0}\right)\,
g_{\alpha\nu}\,
T_{\ell}(i\to\nu)
\;\;\; ,
\label{prodallint}
\end{equation}
where the $g_{\alpha i}$ stand for the relative couplings of each of
the two-meson systems $i$ to a $q\bar{q}$ state of flavor $\alpha$,
and $j_{\ell}$ and $h^{(1)}_{\ell}$ are the spherical Bessel and the
spherical Hankel function of the first kind, respectively.
The interaction radius $r_{0}$ represents
the average distance at which the meson pair emerges from the interaction
region. Furthermore, $\vec{p}_{i}$ is
the relative linear momentum in two-meson channel $i$.
The matrix element $T_{\ell}(\nu\to i)$ of
the scattering amplitude $T$ describes transitions between
channels $\nu$ and $i$ in inelastic meson-meson scattering.

In the form presented in Eq.~(\ref{prodallint}),
the production amplitude consists of a non-resonant term,
given by a Bessel function, and a sum of terms proportional
to matrix elements of the scattering amplitude.
Resonances reside in the latter matrix elements.
Consequently, we have a clear separation of non-resonant
and resonant terms for the production amplitude.
We shall demonstrate in the following that such a separation
is useful.

Expression~(\ref{prodallint}), which was derived in the framework
of the Resonance-Spectrum Expansion (RSE)
\cite{NTTP4}, is a direct link between quark dynamics
and quantities observed in experiment.
However, in its present form the RSE scattering amplitude
does not yet fully correspond to the physical reality,
since it does not account well for
the experimental observation \cite{PLB69p503}
that two-meson channels only couple significantly
to the quark-antiquark propagator in a very limited
range of energies close to threshold.
From harmonic-oscillator (HO) confinement,
which is the basis of the RSE, it is reasonably well understood
why a specific channel does not survive at higher invariant masses.
Namely, it is straightforward to derive \cite{ZPC17p135,ZPC21p291}
that the number of different channels to which the $q\bar{q}$ system couples
grows rapidly for higher excitations.
So for a given probability of quark-pair creation, the fraction that is left
for a specific channel decreases accordingly.

All terms of the production amplitude of Eq.~(\ref{prodallint})
are proportional to three-meson couplings, given by $g_{\alpha\nu}$.
However, the three-meson couplings also enter
the scattering amplitude $T$ itself \cite{AP323p1215}, that is, their
squares. So at higher energies the terms containing the resonances
fall off faster than the non-resonant term,
leading to $q\bar{q}$ resonance peaks that fade away against
background fluctuations of the non-resonant contribution
in production cross sections. In Refs.~\cite{EPL85p61002,ARXIV09044351},
we showed that signs of the $\psi(5S)$ and $\psi(4D)$
charmonium resonances can be observed just above threshold
in the $e^{+}e^{-}\to\Lambda_{c}^{+}\Lambda_{c}^{-}$
cross section measured by the Belle Collaboration \cite{PRL101p172001},
with tentative $\psi(6S)$ and $\psi(5D)$ signals,
at about 400 MeV higher invariant masses,
almost drowned in the background.

By the use of Eq.~(\ref{prodallint})
and the three-meson couplings of Refs.~\cite{ZPC17p135,ZPC21p291},
we may thus understand why it is hard to observe
$q\bar{q}$ resonance peaks in scattering and especially production.
However, from Ref.~\cite{PLB69p503} we conclude that
such a suppression may even be considerably larger,
ranging from a factor $49\pm 25$, determined via the process
$\sigma\left( DD^{\ast}\right)/\sigma\left( D^{\ast}D^{\ast}\right)$,
to a factor $124\pm 100$, obtained from
$\sigma\left( DD\right)/\sigma\left( D^{\ast}D^{\ast}\right)$,
thereby already accounting for the combinatorial factors $PP:PV:VV=1:4:7$
and scaling invariant mass with the RSE HO frequency $\omega$.
Consequently, we must take one step back in order to figure out what
extra ingredient is needed to comply with the experimental evidence.
Here, we focus on two-meson production reactions in $e^{+}e^{-}$ annihilation,
which we assume to take place via the coupling of the $q\bar{q}$ propagator
to the photon.

\begin{figure}[htbp]
\begin{center}
\begin{tabular}{c}
\scalebox{1.0}{\includegraphics{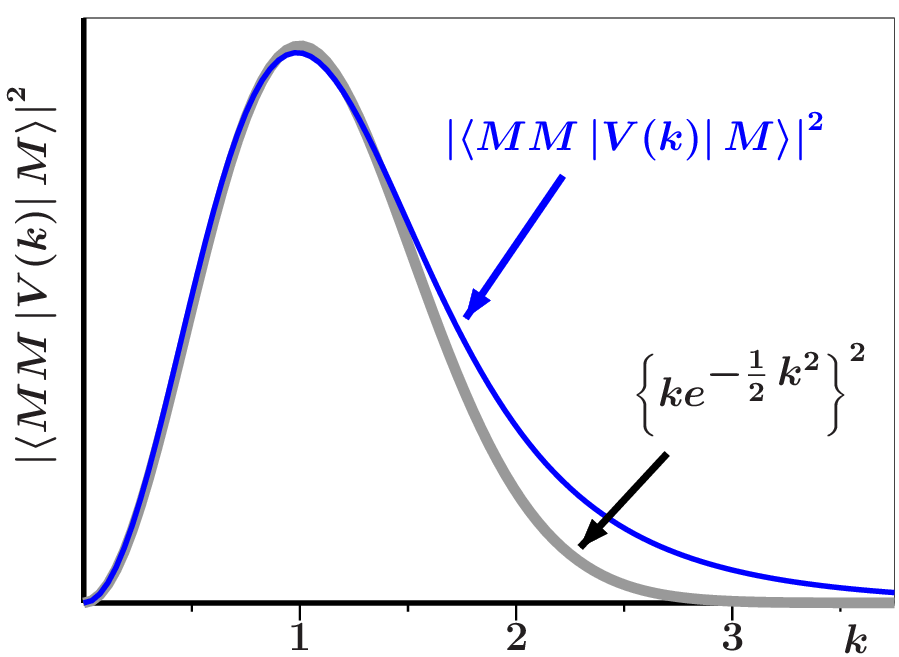}} \\
\end{tabular}\\ [-15pt]
\end{center}
\caption[]{\small The shape of the three-meson vertex strength
$\braket{MM}{V(k)}{M}$ as a function of the relative meson-meson
linear momentum $k$ \lijntje{0}{0}{1},
for $V\to PP$, $V\to PV$ and $V\to VV$
($V=$ vector, $P=$ pseudoscalar)
transitions \cite{ZPC21p291},
assuming HO confinement and $^{3}P_{0}$ quark-pair creation,
and its approximation by a Gaussian shape \lijntje{0.6}{0.6}{0.6}.
The matrix element $\braket{MM}{V(k)}{M}$
and the linear-momentum variable $k$,
are dimensionless here.
}
\label{VtoPP}
\end{figure}

In Refs.~\cite{ZPC17p135,ZPC21p291}, a relation between intermesonic distances
and the strength of the three-meson coupling for configuration space was
derived, assuming HO confinement and $^{3}P_{0}$ quark-pair creation.
However, since for HO distributions
Fourier transforms are obtained by the simple substitution
of coordinate $x$ by linear momentum $k$,
one readily gets the corresponding relation
between the three-meson coupling and relative linear momentum.
In Fig.~\ref{VtoPP}, we depict the shape
of the three-meson vertex intensity \cite{ZPC21p291}
for the cases to be considered here. In the same figure, we show
that this shape, being an approximation itself,
can be further approximated quite well by a Gaussian shape
\cite{JPG36p075002}.

For $e^{-}e^{+}$ annihilation
into hadrons in the charmonium region of invariant masses,
we assume that such processes take place
through the $c\bar{c}$ propagator, and, moreover, that the amplitudes are
dominated by loops of open-charm mesons.
The latter assumption implies that,
even if the final state contains charmless hadrons,
the line shape of its production cross section is to a large extent
proportional to the line shape of open-charm production.

The oscillator states must be decomposed
into two-meson plane waves \cite{PRC77p055206},
which leaves us with the relative linear momentum $p$
as the relevant parameter, rather than the total invariant mass.
In the following we shall demonstrate that,
for simple one-channel reactions, in the absence of inelasticity,
like e.g.\
$D\bar{D}$ production in $e^{+}e^{-}$ annihilation
far below the $D\bar{D}^{\ast}$ threshold,
we may reduce expression (\ref{prodallint})
for the resulting partial $P$-wave cross section
$\sigma_{P}$ to something of the form
\begin{equation}
\sigma_{P}=2\lambda^{2}4\pi\alpha^{2}r_{0}^{2}\abs{
\fnd{p}{\sqrt{s}}\,
pr_{0}\, e^\x{-\frac{1}{2}\left( pr_{0}\right)^{2}}\,
\left\{ j_{1}\left( pr_{0}\right)
+\frac{i}{2}\, h^{(1)}_{1}\left( pr_{0}\right)\,
be^{i\varphi}\,
A\left(\xrm{BW}\right)\right\}}^{2}
\;\;\; ,
\label{bwpbg}
\end{equation}
where, for the case of a reasonably narrow nearby resonance,
such as e.g.\ $\psi (1D)$,
we have simplified the part of Eq.~(\ref{prodallint})
containing the scattering amplitude
to a phase factor multiplying the Breit-Wigner (BW) expression
\begin{equation}
A\left(\xrm{BW}\right) =\fnd{\Gamma /2}{\sqrt{s}-M+i\Gamma /2}
\;\;\; .
\label{bw}
\end{equation}
here, $\sqrt{s}$ and $M-i\Gamma /2$ are
the total invariant mass and the pole position
of the resonance in the complex $E$ plane, respectively.
The dimensionless parameter $\lambda$
represents an overall normalization,
while $b$ adjusts the relative intensities of the resonant
and the non-resonant term.

\begin{figure}[htbp]
\begin{center}
\begin{tabular}{|c|c|}
\hline
\scalebox{0.7}{\includegraphics{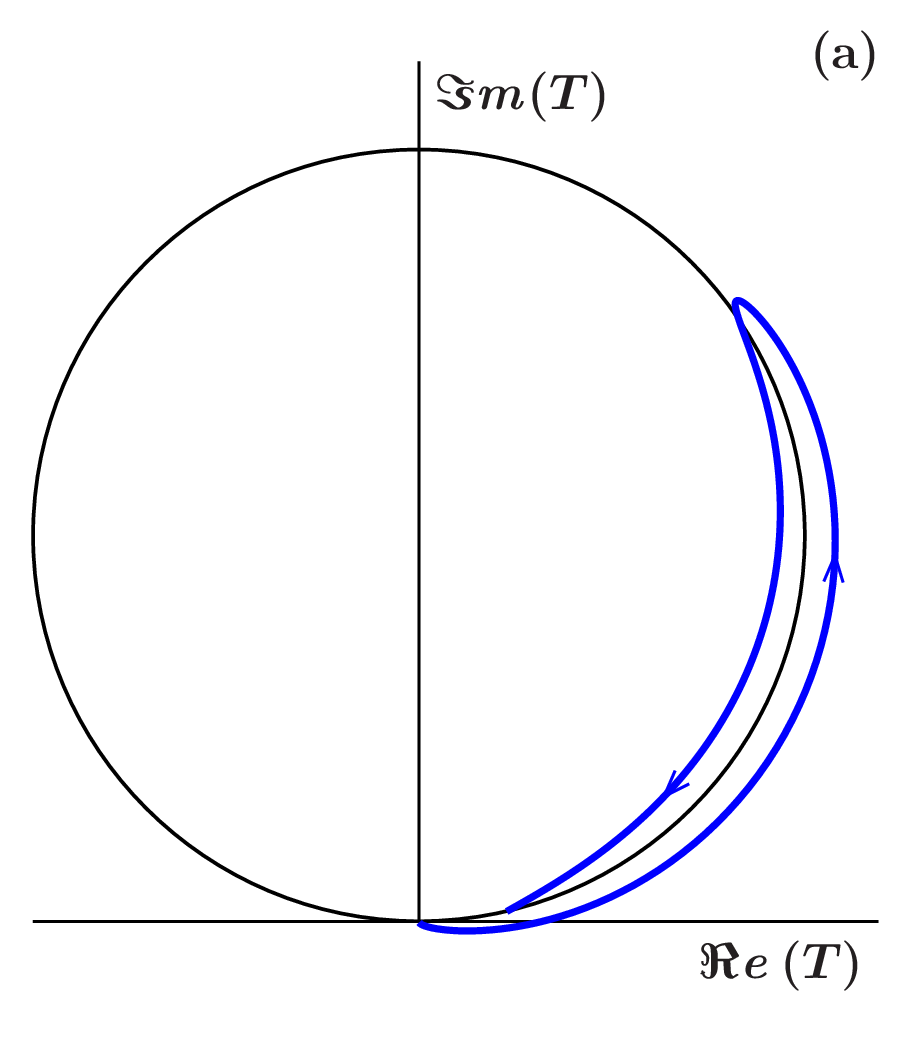}} &
\scalebox{0.7}{\includegraphics{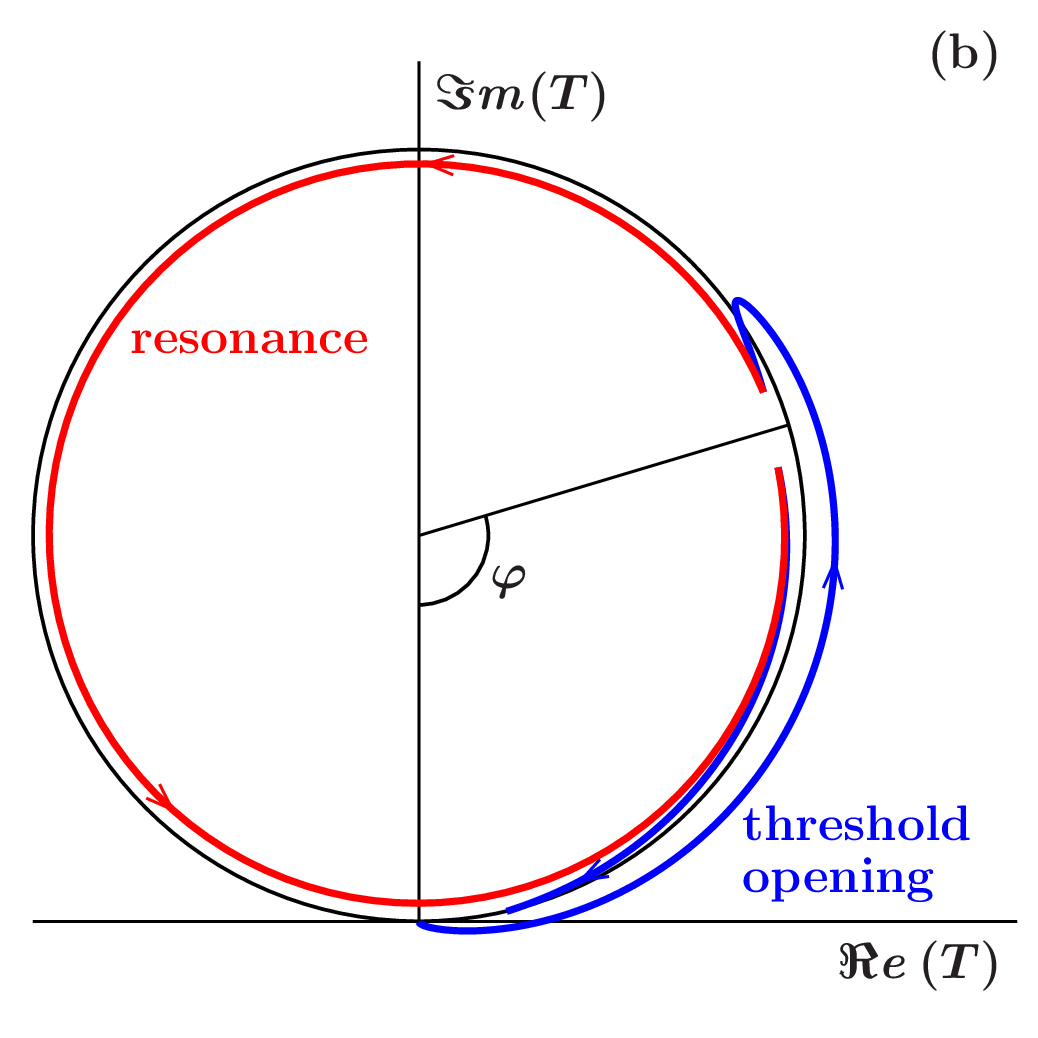}} \\
\hline
\end{tabular}\\ [-15pt]
\end{center}
\caption[]{\small Schematic Argand diagram for the phase motion
of a two-meson production amplitude,
for the case that no resonance is present (a),
and for the case that there also is a resonance
in the same invariant-mass interval (b).
At threshold we assume zero phase.
The arrows indicate the direction of increasing invariant mass.
The angle $\varphi$, representing
the angle of the phase motion of the non-resonant
part of the amplitude at the central resonance mass,
is defined in Eq.~(\ref{bwpbg}).
}
\label{Argand}
\end{figure}

In order to understand well the meaning of the phase
in Eq.~(\ref{bwpbg}), we depict in Fig.~\ref{Argand}
an elastic-scattering equivalent of Eq.~(\ref{bwpbg}).
Of course, there is no elastic-unitarity requirement in production.
Hence, unambiguous definitions of production phases
cannot be made.
Moreover, disentangling non-resonant signals
and resonance amplitudes is model dependent,
even in the most naive approaches.
Here, we propose to use our model \cite{AP323p1215}
for which the scattering amplitude $T$ contains
the quark-antiquark propagator spectrum from the start,
and for which some of the resulting resonances can be approximated
by BW amplitudes.

In Fig.~\ref{Argand}a, we sketch the phase motion of a non-resonant signal,
as an illustration. At threshold we assume a vanishing phase.
The phase grows with increasing invariant mass,
up to a certain maximum, after which it decreases.
The maximum may be well over $\varphi =90^{\circ}$,
but this does not necessarily imply that
we are dealing with a resonance (cf.\ Levinson's theorem).
The phase motion of a resonance,
riding on top of the falling slope of the non-resonant signal,
is indicated in Fig.~\ref{Argand}b.
While the phase of the non-resonant contribution moves on,
the phase of the resonance makes a full $180^{\circ}$ turn,
to end up back on the slope of the non-resonant contribution
at some higher invariant mass.
The average phase $\varphi$ of the non-resonant contribution in
Eq.~(\ref{bwpbg}) is assumed to be a constant.

\section{The parameter \bm{r_{0}}}
\label{production}

The parameter $r_{0}$,
which was introduced in Ref.~\cite{PRD21p772}
in order to allow for a quantity representing
the average distance of quark-pair creation,
requires renewed attention here.

In Fig.~\ref{VtoPP} we have depicted the shape of
the $^{3}P_{0}$ coupling as a function of
the two-meson relative linear momentum $p$.
We find that it vanishes at $p=0$,
rises to a maximum, and then drops fast
for increasing values of $p$.
This behavior of the three-meson vertex intensity
as a function of two-meson relative linear momentum
agrees with the experimental observation obtained
with the SLAC/LBL magnetic detector at SPEAR \cite{PLB69p503}.
The results suggest that at higher invariant masses the branching
fraction of a specific two-meson channel vanishes rapidly.
However, the rate at which the branching fractions decrease
still remains to be studied.

As an example,
let us concentrate on the reaction $e^{+}e^{-}\to B\bar{B}$.
In the RSE, the $B\bar{B}$ threshold at 10.56 GeV comes out
far above the HO ground state at 9.72 GeV
and the first radial excitation at 10.10 GeV,
and also above the second radial excitation at 10.48 GeV.
Note that the radial level spacings are equal
to $2\omega =0.38$~GeV in the RSE.

The three-meson vertex intensities for these HO levels
can be determined with the formalism developed in
Ref.~\cite{ZPC21p291}.
The total $B\bar{B}$ branching fraction is found to be
roughly 2.8\% at the HO ground state,
but only about 0.4\% at the second radial HO excitation.
The main competition for $B\bar{B}$ formation
above the $B\bar{B}$ threshold stems from the $BB^{\ast}$ channel.
This competing channel will start to dominate
for invariant masses closer to the $BB^{\ast}$ threshold at 10.60 GeV.
Above the $BB^{\ast}$ threshold, we suppose
that $B\bar{B}$ formation rapidly vanishes.

The above implies that at the $BB^{\ast}$ threshold
the coupling of $B\bar{B}$ to the $b\bar{b}$ propagator
must have decreased significantly.
\begin{figure}[htbp]
\begin{center}
\begin{tabular}{c}
\scalebox{0.6}{\includegraphics{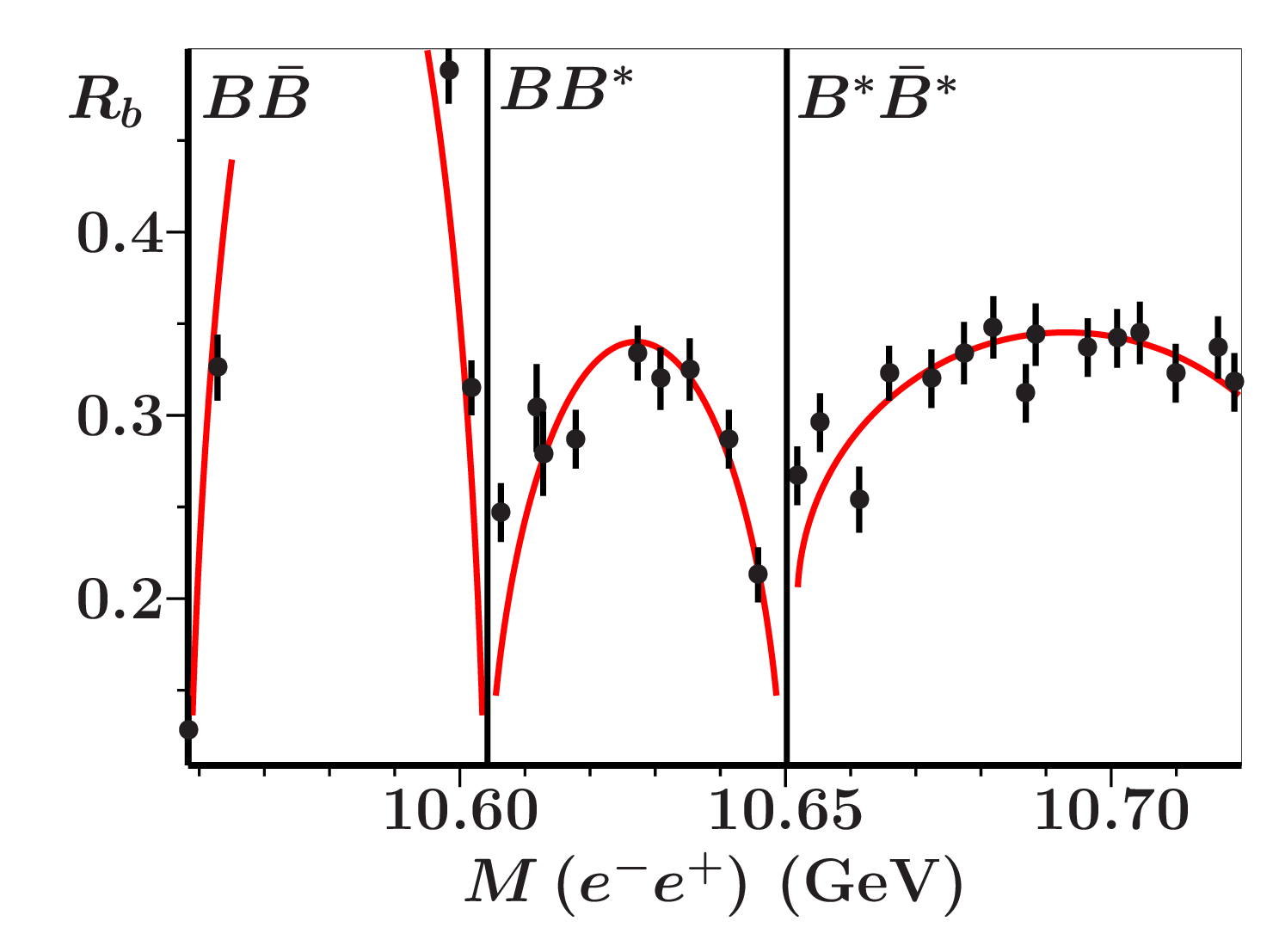}}\\ [-15pt]
\end{tabular}
\end{center}
\caption[]{\small
Experimental data for the process $e^{+}e^{-}\to b\bar{b}$
obtained by the BaBar Collaboration
in Ref.~\cite{PRL102p012001}.
The vertical lines indicate the $BB^{\ast}$ and $B^{\ast}B^{\ast}$
thresholds, as indicated in the figure.
The eye-guiding lines reflect our interpretation
of the data, and do not represent fits to the data.}
\label{morebabar}
\end{figure}
This is indeed confirmed by Fig.~\ref{morebabar},
in which we show data for the process $e^{+}e^{-}\to b\bar{b}$,
measured and analysed by the BaBar Collaboration \cite{PRL102p012001}.
As also remarked in their paper,
the large statistics and the small energy steps of the
scan make it possible to clearly observe the two dips
at the opening of the thresholds corresponding
to the $B\bar{B}^{\ast}+\bar{B}B^{\ast}$
and $B^{\ast}\bar{B}^{\ast}$ channels.

We have cut off the huge peak at 10.58 GeV,
in order to concentrate better on
the details of the other two enhancements,
at 10.63 GeV and 10.69 GeV, respectively.
Near the $BB^{\ast}$ threshold, we thus observe that
the $B\bar{B}$ signal rapidly
vanishes for increasing invariant mass,
whereas the $BB^{\ast}$ signal behaves the opposite way, i.e.,
it grows fast for increasing invariant mass
just above threshold.
At the $B^{\ast}B^{\ast}$ threshold, this phenomenon is repeated,
now with respect to the $BB^{\ast}$ signal.

So we conclude that $B\bar{B}$ production in $e^{+}e^{-}$ annihilation
can mainly be observed within the invariant-mass window
delimited by the $B\bar{B}$ and $BB^{\ast}$ thresholds.
Similarly, $BB^{\ast}$ production has an equally wide window formed
by the $BB^{\ast}$ and $B^{\ast}B^{\ast}$ thresholds.
The window for $B^{\ast}B^{\ast}$ production is somewhat wider,
since the next threshold concerns the $B_{s}B_{s}$ channel,
which lies considerably higher. Therefore, the enhancement peaking at
about 10.69~GeV is broader than the ones at 10.58 and 10.63~GeV.

From Fig.~\ref{VtoPP}, we see that the $b\bar{b}\to B\bar{B}$
form factor closes its window at about $k=2.5$. As a result of the
above discussion and considering the structure of the production amplitude
in Eq.~(\ref{prodallint}), this implies that the dimensionless quantity
defined by
\begin{eqnarray}
\lefteqn{p_{B\bar{B}}^{2}
\left(\xrm{at}\;\; M\left( B\bar{B}\right)
=M_{B^{\ast}}+M_{B}\right)\,
r_{0,B\bar{B}}^{2}=}
\nonumber\\ [10pt] & &
=\left\{\frac{1}{4}\left( M_{B^{\ast}}+M_{B}\right)^{2}
-M_{B}^{2}\right\}\, r_{0,B\bar{B}}^{2}
=\frac{1}{4}\left( M_{B^{\ast}}^{2}-M_{B}^{2}\right)
\left\{ 1+\fnd{2M_{B}}{M_{B^{\ast}}+M_{B}}\right\}\, r_{0,B\bar{B}}^{2}
\label{patBBstar}
\end{eqnarray}
must be equal to about
$2.5^{2}\approx \left( 0.5\;\xrm{GeV fm}/\hbar c\right)^{2}$.

Owing to quadratic meson-mass relations,
the quantity $M_{V}^{2}-M_{P}^{2}$
($V=$ vector, $P=$ pseudoscalar meson)
seems to be almost a constant of nature for the cases we shall consider here,
namely $0.488\pm 0.007$ for $V=B^{\ast}$ and $P=B$,
$0.550\pm 0.001$ for neutral and $0.546\pm 0.002$ for charged
$D^{(\ast )}$ mesons,
and $0.5513\pm 0.0005$ for $K^{(\ast )\pm}$.
Furthermore, for heavy quarks, the expression between
braces on the righthand side
of Eq.~(\ref{patBBstar}) approximately equals 2.
This suggests that we could set out by assuming
that the whole quantity in Eq.~(\ref{patBBstar}) is a constant.
In the following we shall show that indeed the constant
$p_{P\bar{P}}
\left(\xrm{at}\;\; M=M_{V}+M_{P}\right)\,
r_{0,P\bar{P}}=2.5$
leads to good results for $P=D,B$ ($V=D^*,B^*$),
and even for $P=\Lambda_{c}$ ($V=\Sigma_{c}$),
though for $P=K$ ($V=K^*$) we need a somewhat smaller value for this constant.

In view of the above, we find for the $r_{0}$ parameter
in the $c\bar{c}$ and $b\bar{b}$ sectors
considerably larger values than we had assumed in earlier work.
Although this is not entirely unwelcome, as it may help cure the somewhat
too small widths of $c\bar{c}$ and $b\bar{b}$ resonances we obtained before,
it seems to break, to some extent,
the rigorous flavor independence
we advertised in the past.
A more detailed discussion of how to precisely
interpret $r_0$ is required.
The problems lies in the fact that the valence particles involved
in the decay process change their character and
in general their masses, that is,
an original $q\bar{q}$ pair transforms,
in the course of the decay process, into a pair of mesons.
In a coordinate-space description,
which was the original formulation of the model
\cite{PRD21p772,PRD27p1527},
it turned out necessary to assume
that the $\vec{r}$ coordinate of the valence $q\bar{q}$ pair
is the same as for the emerging meson pair,
in order to avoid nonlocalities in the equations.
This approximation is reasonable when both the decaying meson
and the decay products are composed of light quarks,
so that all mesons have comparable sizes.
However, when the original meson is a heavy quarkonium,
which is a much more compact system
especially in the bottomonium case,
the main decay products are mesons with open beauty (or charm),
which are clearly much more extended objects
owing to the presence of a light quark.
In these situations, the local approximation is obviously not ideal,
but the tendency for $r_0$ will be to adjust itself
chiefly to the kinematics of the outgoing meson pair,
especially when data over a wide $p$ range are analysed.
Possible improvements will be left to future work.
One alternative is the employment of more extended transition
potentials \cite{ZPC21p291}, so as to allow for more detail in the $r_0$
dependence of the wave functions.
Another possibility is to work in a covariant momentum-space framework
\cite{AIPCP660p367}, in order to be able to deal with nonlocalities.

In view of these considerations, it seems most appropriate to consider
$r_{0}$, which appears in the Bessel and Hankel functions of
Eq.~(\ref{prodallint}),
a parameter exclusively governed
by the meson-meson dynamics and kinematics,
at least for the purpose of the present analysis.
The mild deviations from exact flavor independence of the meson spectra,
resulting from this approach, are irrelevant for this work,
and would  generally have only minor effects
on the central resonance positions predicted by the RSE.

In the following we compare $e^{+}e^{-}$ annihilation data
above the open-strangeness threshold
for $K^{+}K^{-}$ production \cite{PLB107p297,PLB99p257},
just above the open-charm threshold
for inclusive hadron production \cite{ARXIV08070494},
just above the open-beauty threshold
for inclusive hadron production \cite{PRD72p032005}
and exclusive $b\bar{b}$ production \cite{PRL102p012001},
and just above the open-charm baryon-pair threshold
for $\Lambda_{c}^{+}\Lambda_{c}^{-}$ production \cite{PRL101p172001}.

The production cross sections of hadron pairs
are expressed in nanobarns in the above experimental analyses.
Hence, we must scale expression (\ref{bwpbg})
with the size of the available window for production
and by the appropriate conversion factors.
Thereto, by the use of the quantity given in Eq.~(\ref{patBBstar})
and its equivalent for the other processes studied here,
we perform in expression (\ref{bwpbg}) the substitution
\begin{equation}
\lambda^{2}r_{0}^{2}\to\frac{1}{2}{\lambda '}^{2}
\fndrs{5pt}{r_{0}^{2}\times 10^{7}\;\xrm{nb/fm$^{2}$}}
{-5pt}{\sqrt{2.5^{2}+\left( m_{h}r_{0}\right)^{2}}-m_{h}r_{0}}
\;\;\; ,
\label{scaling}
\end{equation}
in order to facilitate direct comparison of Eq.~(\ref{bwpbg})
with experimental results.
For $m_{h}$ given by $m_{D}$, $m_{B}$, $m_{\Lambda_{c}}$, and $m_{K}$,
we next adjust the shape in Eq.~(\ref{bwpbg})
to the shapes of the experimental cross sections,
by choosing for $\lambda '$ the values 0.95, 1.03, 1.02,
and 0.49 (only $K^{+}K^{-}$ has been measured),
respectively.

\subsection{The reaction \bm{e^{+}e^{-}\to D\bar{D}}}

The reaction of $e^{+}e^{-}$ annihilation
into open-charm pairs can be observed
at and above the $D\bar{D}$ threshold.
We assume here that the reaction takes place via a photon
and the $c\bar{c}$ propagator,
through the creation of a light $q\bar{q}$ pair.
However, as we have discussed previously,
many competing configurations may be formed,
increasing in number for higher invariant masses.
Furthermore, it seems we may conclude from experiment
that stable open-charm hadrons have more probability to be
formed near threshold, where kinetic energy is almost zero.
Hence, if it were not for phase space and the centrifugal barrier,
$D\bar{D}$ pairs would be produced most likely just above threshold.

For total invariant mass below, but close to the $DD^{\ast}$ threshold,
we assume that the probability of $D\bar{D}$ creation
is already reduced because of the non-vanishing probability of creating a
virtual $DD^{\ast}$ pair. Just above the $DD^{\ast}$ threshold,
$D\bar{D}$ creation decreases rapidly.
The expected corresponding non-resonant contribution
of the production amplitude should thus exhibit this feature.
In this respect, an important observation was published
by the BES Collaboration in Ref.~\cite{ARXIV08070494}.
To our knowledge, the BES Collaboration was the first to discover
that the $\psi (3770)$ cross section is built up
by two different amplitudes, viz.\ a relatively broad signal and a
rather narrow $c\bar{c}$ resonance.
For the narrow resonance, which probably corresponds to
the well-established $\psi (1D)(3770)$,
the BES Collaboration measured a central resonance position
of $3781.0\pm 1.3\pm 0.5$ MeV
and a width of $19.3\pm 3.1\pm 0.1$ MeV (their solution 2).
If the latter parameters are indeed confirmed,
it would be yet another observation
of a quark-antiquark resonance width
that is very different from the world average
($83.9\pm 2.4$ MeV \cite{PLB667p1} in this case),
after a similar result was obtained by the BaBar Collaboration
in Ref.~\cite{PRL102p012001}, for $b\bar{b}$ resonances.
Concerning the broader structure, the BES Collaboration indicated, for
their solution 2, a central resonance position of $3762.6\pm 11.8\pm 0.5$ MeV
and a width of $49.9\pm 32.1\pm 0.1$ MeV.
The signal significance for the new enhancement is $7.6\sigma$
(solution 2). It is has been explained as a possible diresonance
\cite{PRD78p116014} or heavy molecular state \cite{ARXIV08080073}.

Furthermore, in the latter BES publication, the existence of conflicting
results with respect to the branching fraction
for non-$D\bar{D}$ hadronic decays of the $\psi (1D)(3770)$ was emphasized.
On the one hand, the total branching fraction
for exclusive non-$D\bar{D}$ modes has been measured
to be less than 2\% \cite{PLB605p63,PRD74p012005}.
But on the other hand, for inclusive non-$D\bar{D}$ decay modes,
values of about 15\% have been found \cite{PRD76p122002,PLB659p74}.
According to the BES Collaboration, this apparent discrepancy may partially
be due to the assumption that the line shape above the $D\bar{D}$ threshold
is the result of one simple resonance.
Here, we shall show that the broader structure is most likely
caused by the non-resonant contribution
to the production amplitude given in Eq.~(\ref{bwpbg}),
thus lending further support to the idea that
the $\psi (1D)(3770)$ enhancement consists of two distinct signals.

We mentioned in the above the recent findings of the BaBar Collaboration
with respect to the resonance parameters of $b\bar{b}$ states.
The corresponding measured line shapes actually show a similar behavior,
namely a broad structure with a narrow resonance on top.
Nevertheless, the BaBar Collaboration did not conclude
that the broad structures are due to additional resonances
in the $b\bar{b}$ sector.
As we shall demonstrate in the following,
non-resonant contributions can indeed be large,
sometimes even much larger than the $q\bar{q}$ resonances
in their vicinity.

\begin{figure}[htbp]
\begin{center}
\begin{tabular}{c}
\scalebox{1.0}{\includegraphics{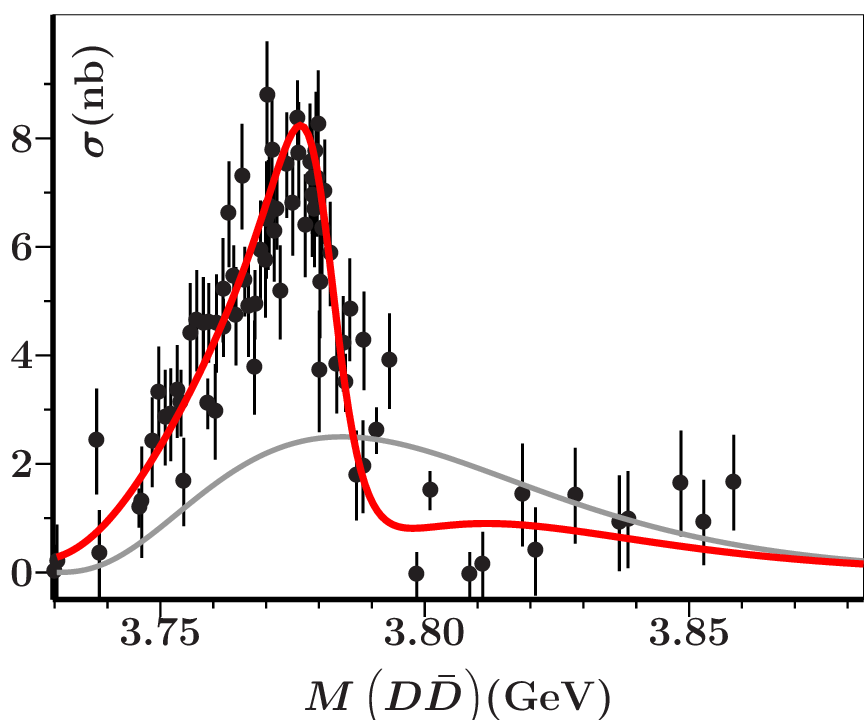}}\\ [-15pt]
\end{tabular}
\end{center}
\caption[]{\small
Comparison of experimental data obtained by the BES Collaboration
\cite{ARXIV08070494}
and the cross section resulting from Eq.~(\ref{bwpbg}) \lijntje{1}{0}{0}
for $r_{0}=0.962$ fm and $b=1.33$.
The non-resonant contribution \lijntje{0.6}{0.6}{0.6}
dominates for larger relative $D\bar{D}$ linear momentum.
The BW parameters of the $c\bar{c}$ resonance are
$M\left(\psi (1D)\right) =3.781$ GeV and
$\Gamma\left(\psi (1D)\right) =17$ MeV.
}
\label{bes}
\end{figure}

If we assume the above-mentioned value of 2.5
for the quantity in Eq.~(\ref{patBBstar}), we get $r_{0}=0.962$~fm.
A comparison of the resulting theoretical curve,
by the use of Eq.~(\ref{bwpbg}),
together with the data of the BES Collaboration,
is shown in Fig.~\ref{bes}.
The BW parameters of the $\psi (3770)$
used for the theoretical curve of Fig.~\ref{bes},
are collected in Table.~\ref{psi3770}
and compared to the corresponding PDG parameters
of those of the BES Collaboration.
\begin{table}[htbp]
\begin{center}
\begin{tabular}{||c||ccc||}
\hline\hline & & & \\ [-7pt]
source & Mass (GeV) & Width (MeV) & phase (rad)\\
& & & \\ [-7pt]
\hline & & & \\ [-7pt]
Fig.~\ref{bes} & 3.781 & 17 & $0.75\pi$\\ [10pt]
BES \cite{ARXIV08070494} & $3.7810\pm 1.3\pm 0.5$ &
$19.3\pm 3.1\pm 0.1$ & $(0.88\pm 1.86\pm 0.03)\pi$\\ [10pt]
PDG \cite{PLB667p1} & $3.7700\pm 0.0024$ & $83.9\pm 2.4$ & \\ [10pt]
\hline\hline
\end{tabular}
\end{center}
\caption[]{\small Parameters used for the solid line
in Fig.~\ref{bes},
compared to the corresponding parameters obtained by
the BES Collaboration \cite{ARXIV08070494},
and to world averages \cite{PLB667p1}.}
\label{psi3770}
\end{table}

\subsection{The reaction \bm{e^{+}e^{-}\to B\bar{B}}}

The higher excitations of the beautonium vector states,
discovered more than two decades ago,
are still today a puzzling topic of intensive research.
In Refs.~\cite{PRL54p377} and \cite{PRL54p381},
the CUSB and CLEO Collaborations, respectively, displayed
the first results for the invariant-mass spectra of the
$R\left(\sigma_\xrm{\scriptsize had}/\sigma_{\mu\mu}\right)$
ratio above the open-beauty threshold.

The data of Ref.~\cite{PRL54p377} were observed with
the CUSB calorimetric detector operating at CESR (Cornell).
The experimental analysis resulted in evidence for structures at
$10577.4\pm 1$ MeV,
$10845\pm 20$ MeV, and
$11.02\pm 0.03$ GeV,
with total hadronic widths of
$25\pm 2.5$ MeV,
$110\pm 15$ MeV, and
$90\pm 20$ MeV, respectively.
Structures at about 10.68 and 11.2 GeV were not included
in the analysis of the CUSB Collaboration.

The data of Ref.~\cite{PRL54p381} were obtained from
the CLEO magnetic detector, also operating at CESR.
The experimental analysis resulted in evidence for structures at
$10577.5\pm 0.7\pm 4$ MeV,
$10684\pm 10\pm 8$ MeV,
$10868\pm 6\pm 5$ MeV, and
$11019\pm 5\pm 5$ MeV,
with total hadronic widths of
$20\pm 2\pm 4$ MeV,
$131\pm 27\pm 23$ MeV,
$112\pm 17\pm 23$ MeV, and
$61\pm 13\pm 22$ MeV, respectively.
A structure at about 11.2 GeV was not included
in the analysis of the CLEO Collaboration.

The enhancement at 10.580 GeV was extensively studied
in a more recent publication
\cite{PRD72p032005} of the BaBar Collaboration.
Data were collected with the BaBar detector at the PEP-II storage ring
of Stanford Linear Accelerator Center.
The experimental analysis yielded
$10579.3\pm 0.4\pm 1.2$ MeV and $20.7\pm 1.6\pm 2.5$ MeV
for the central mass and the total (hadronic) width, respectively,
which in in fair agreement with the above results
of the CUSB and CLEO Collaborations.

A very recent release of the BaBar Collaboration
\cite{PRL102p012001}
is devoted to two more resonances in the $b\bar{b}$ spectrum,
also using data obtained by the BaBar detector at the PEP-II storage ring,
resulting in the BW parameters
$10876\pm 2$ MeV (mass) and $43\pm 4$ (width)
for the $\Upsilon (10860)$, and
$10996\pm 2$ MeV (mass) and $37\pm 3$ (width)
for the $\Upsilon (11020)$,
which differ substantially, in particular for the widths,
from the above results
of the CUSB and CLEO Collaborations,
and also from the world-average values \cite{PLB667p1}
$10865\pm 8$ MeV (mass) and $110\pm 13$ (width)
for the $\Upsilon (10860)$ and
$11019\pm 8$ MeV (mass) and $79\pm 16$ (width)
for the $\Upsilon (11020)$.
Of course, such discrepancies call for further study.
However, in this work we shall only focus
on the reaction $e^{+}e^{-}\to B\bar{B}$
for total invariant masses below the $BB^{\ast}$ threshold.

\begin{figure}[htbp]
\begin{center}
\begin{tabular}{cc}
\scalebox{0.8}{\includegraphics{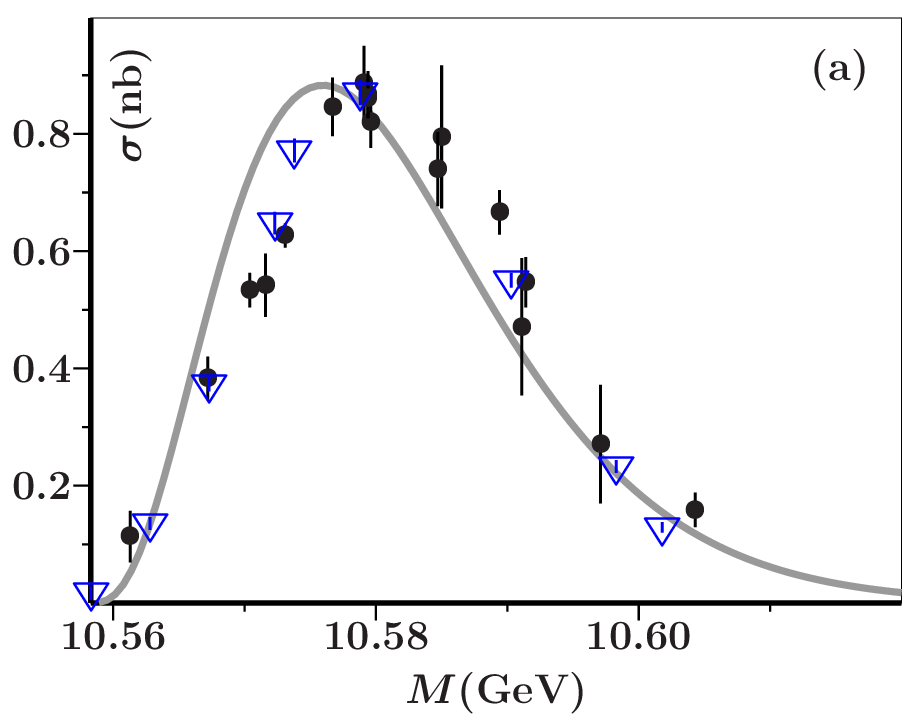}} &
\scalebox{0.8}{\includegraphics{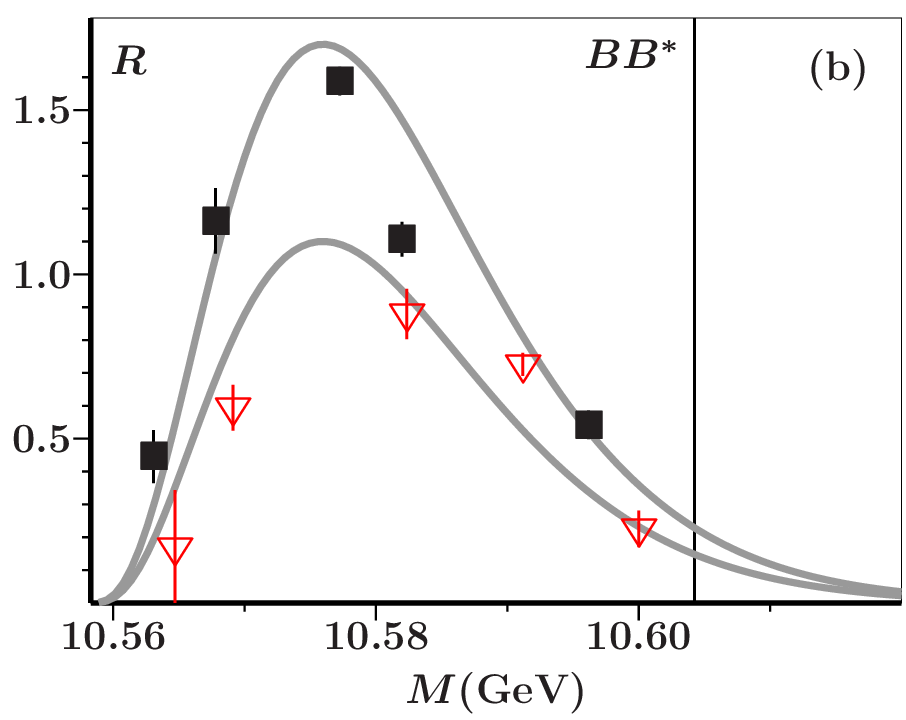}}\\ [-15pt]
\end{tabular}
\end{center}
\caption[]{\small
Comparison of experimental data taken from
(a) Refs.~\cite{PRD72p032005} ($\bullet$) and
\cite{PRL102p012001} ({\color{blue}$\bigtriangledown$}),
and (b) Refs.~\cite{PRL54p377} ({\color{red}$\bigtriangledown$})
and \cite{PRL54p381} ($\blacksquare$),
with the line shape resulting from Eq.~(\ref{bwpbg})
for $r_{0}=1.014$ fm.
The vertical line in (b)
indicates the position of the $BB^{\ast}$ threshold.
}
\label{up10580}
\end{figure}

The data of the BaBar Collaboration of Ref.~\cite{PRD72p032005}
consist of three different scans of the visible cross section
for $b\bar{b}$ production, with a background of other hadronic
processes in $e^{+}e^{-}$ annihilation.
The three scans differ in the amount of probable background,
while the first scan also has its peak shifted.
Here, we subtract backgrounds of
0.82 nb, 0.91 nb, and 0.99 nb, respectively, from the data of
scans I, II and III.
Moreover, we shift the data of scan I by 4 MeV.
The shape of the resulting signal of 17 data points above
the $B\bar{B}$ threshold is compared to the data
of the BaBar Collaboration \cite{PRL102p012001}.
For the latter data, which show the $R_{b}$ ratio
for all $e^{+}e^{-}$ annihilation processes
containing $b$ quarks, we assume
a background of $R_{b}=0.1$, to account for
those reactions that do not contain $B\bar{B}$ pairs.
Fig.~\ref{up10580}a shows the resulting line shape
of the enhancement at 10.58 GeV.

The data of the CUSB \cite{PRL54p377}
and CLEO \cite{PRL54p381} Collaborations
are shown in Fig.~\ref{up10580}b.
The former data are as measured, but subtracted by $R$(visible) $=2.3$
in order to account for background, as suggested by CUSB themselves.
Similarly, the latter data are as measured,
but subtracted by $R=4.57$ to account for background,
as suggested by CLEO.
We have not put the CUSB and CLEO data in one figure together
with the BaBar ones, since it is not clear to us how to normalize
each separate data set so as to make sense out of a joint presentation.
Moreover, in Fig.~\ref{up10580}b the CUSB and CLEO data are
treated as two different sets, each compared separately to
the shape resulting from Eq.~(\ref{bwpbg}).

The theoretical curves in Fig.~\ref{up10580} correspond to
the non-resonant contribution in Eq.~(\ref{bwpbg}).
In Ref.~\cite{PLB667p1}, this enhancement is classified as a $b\bar{b}$
resonance, under $\Upsilon (10580)$. However, in view of our results, we do
not believe this enhancement to represent a resonance.
In fact, we expect \cite{PRD21p772} the corresponding $4^{\; 3}S_{1}$
$b\bar{b}$ resonance to lie at roughly 10.77 GeV.
A better candidate for this $b\bar{b}$ resonance
is the state found by the CLEO Collaboration \cite{PRL54p381}
at $10684\pm 10\pm 8$ MeV.
It is unclear to us why this resonance, observed back in 1985,
was classified as a $b\bar{b}g$ hybrid state by the CLEO Collaboration,
and, subsequently, not even mentioned in the PDG tables.

The enhancement at 10.58 GeV suggests
an accumulation of $B\bar{B}$ pairs in this invariant-mass region.
Therefore, a description in terms of a wave function
with a dominant $B\bar{B}$ component appears to be more adequate
than assuming a pole in the scattering amplitude
due to a supposed underlying $b\bar{b}$ state.
%\clearpage

\subsection{The reaction \bm{e^{+}e^{-}\to\Lambda_{c}^{+}\Lambda_{c}^{-}}}

In Ref.~\cite{PRL101p172001}, the Belle Collaboration announced
the observation of a near-threshold enhancement,
by studying the $e^{+}e^{-}\to\Lambda_{c}^{+}\Lambda_{c}^{-}$ cross section.
The experimental analysis resulted in a mass and width
for this enhancement of $M=(4634^{+8}_{-7})$(stat.)$^{+5}_{-8}$(sys.) MeV
and $\Gamma_{\mathrm{tot}}=92^{+40}_{-24}$(stat.)$^{+10}_{-21}$(sys.) MeV,
respectively \cite{PRL101p172001}, with a significance of $8.8$ $\sigma$.
An intriguing aspect of this experimental observation
is that the main signal lies close to
the $\Lambda_{c}^{+}\Lambda_{c}^{-}$ threshold,
making an understanding of this structure a highly topical issue.

\begin{figure}[htbp]
\begin{center}
\begin{tabular}{c}
\scalebox{1.0}{\includegraphics{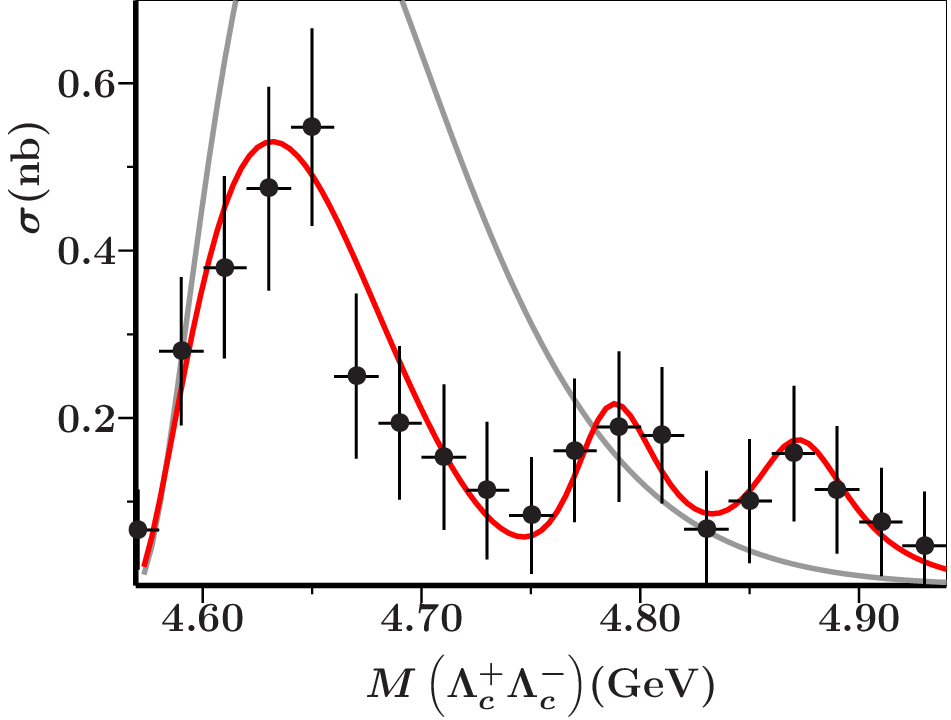}}\\ [-15pt]
\end{tabular}
\end{center}
\caption[]{\small
Comparison of the experimental cross section
for the reaction
$e^{+}e^{-}\to\Lambda_{c}^{+}\Lambda_{c}^{-}$,
obtained by the Belle Collaboration \cite{PRL101p172001},
and the line shape resulting from
the $S$-wave equivalent of Eq.~(\ref{bwpbg})
\lijntje{1}{0}{0}
for $r_{0}=0.962$ fm
(see also Ref.~\cite{JPG36p075002}).
The non-resonant contribution \lijntje{0.6}{0.6}{0.6}
is relatively large.
The BW parameters for the two $c\bar{c}$ resonances
\cite{EPL85p61002} are
$M\left(\psi (5S)\right) =4.784$ GeV,
$\Gamma\left(\psi (5S)\right) =55$ MeV, and
$M\left(\psi (4D)\right) =4.871$ GeV,
$\Gamma\left(\psi (4D)\right) =60$ MeV.
}
\label{LcLc}
\end{figure}

An initial study of the Belle cross section
was done in Ref.~\cite{EPL85p61002}.
Here, we apply Eq.~(\ref{bwpbg}), but assume an $S$
rather than a $P$ wave,
by choosing the appropriate Bessel and Hankel functions.
The result is shown in Fig.~\ref{LcLc},
where, besides reproducing the enhancement at threshold,
we also spot two new charmonium states, identified as
the $\psi (5S)$ and $\psi (4D)$.
We get 4.784 GeV and 4.871 GeV
for the central resonance masses,
and 55 MeV and 60 MeV for the widths, respectively.
Furthermore, the curves in Fig.~\ref{LcLc} clearly show
that the presence of the $\psi (5S)$ and $\psi (4D)$ $c\bar{c}$ resonances
strongly affects the production of
$\Lambda_{c}^{+}\Lambda_{c}^{-}$ near threshold.
The latter shape will be further influenced by the $\psi (3D)$
$c\bar{c}$ resonance around 4.53 GeV
\cite{PRD21p772,PRD32p189,EPL85p61002}, which still needs confirmation.
However, this state cannot be parametrized with a simple
BW structure, which does not have the right threshold behavior.
Nevertheless, the full RSE amplitude, without the approximation of resonances
by BWs as employed in the present analysis, may be used for the study of
such effects. This will be done in future work.

Modeling the the $\Lambda_{c}^{+}\Lambda_{c}^{-}$
enhancement could of course be done by considering
a full wave function with all possible components, viz.\
$c\bar{c}$ states, charmed-meson pairs, and charmed-baryon pairs.
Such a wave function will have a large
$\Lambda_{c}^{+}\Lambda_{c}^{-}$ component under the enhancement.
Nevertheless, it will not give rise to a resonance pole
in the full coupled-channel scattering amplitude,
unless, accidentally, there is a dynamically generated
resonance pole in this invariant-mass region.
Now, in view of the large non-resonant contribution
to the $\Lambda_{c}^{+}\Lambda_{c}^{-}$ enhancement,
we do not consider the occurrence here of a dynamically generated
resonance very likely,
which is in line with a similar conclusion of D.~V.~Bugg
\cite{JPG36p075002}.
In our opinion, it certainly does not represent
a state of the $J^{PC}=1^{--}$ $c\bar{c}$ spectrum
as advocated in Refs.~\cite{PRD78p114033,PRD79p094004}.

\subsection{The reaction \bm{e^{+}e^{-}\to K^{+}K^{-}}}

\begin{figure}[htbp]
\begin{center}
\begin{tabular}{cc}
\scalebox{0.85}{\includegraphics{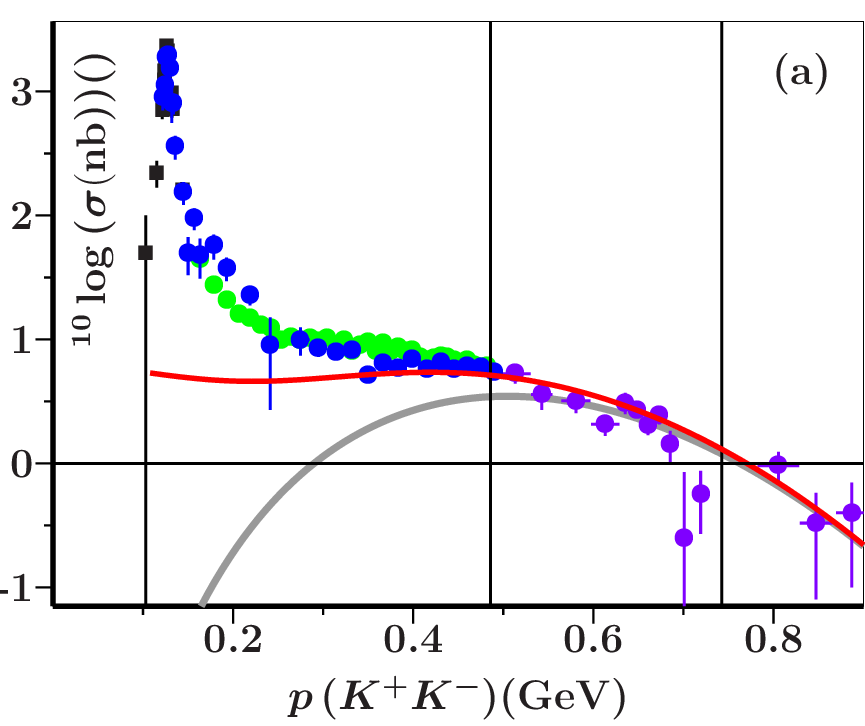}} &
\scalebox{0.85}{\includegraphics{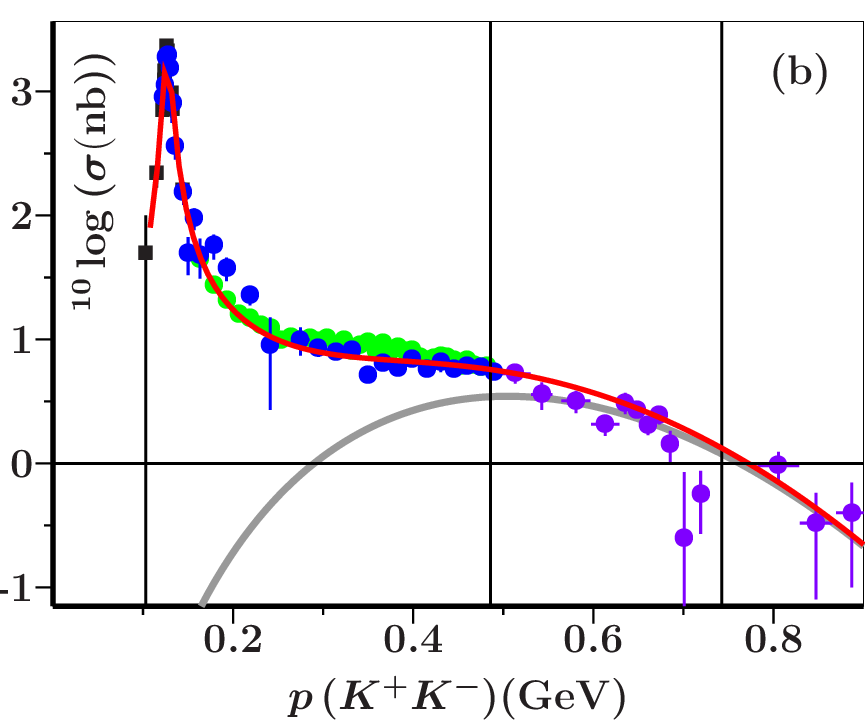}}\\
\scalebox{0.85}{\includegraphics{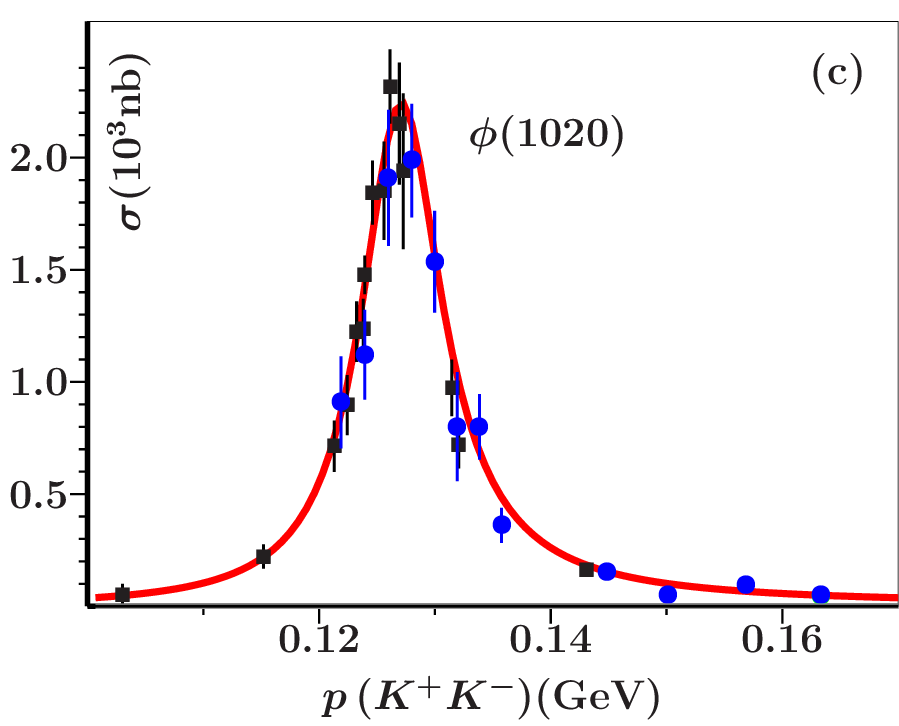}} &
\scalebox{0.85}{\includegraphics{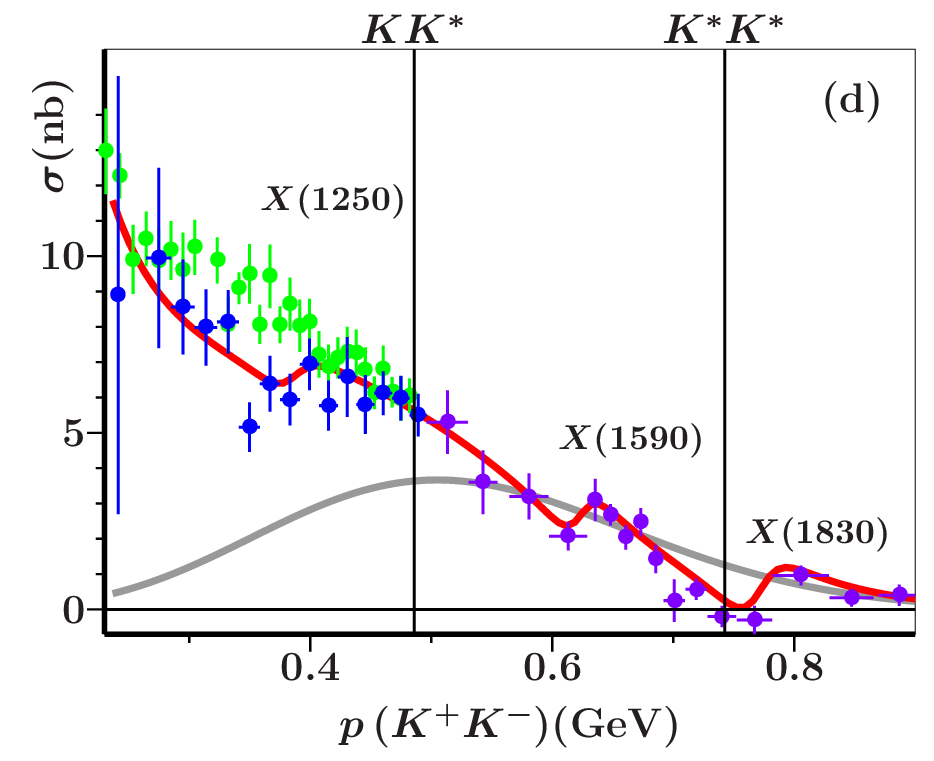}}\\ [-15pt]
\end{tabular}
\end{center}
\caption[]{\small
Comparison of combined experimental data taken from
Refs.~\cite{PLB99p257} ({\color{pink}$\bullet$}),
\cite{PLB107p297} ({\color{blue}$\bullet$}),
\cite{PLB364p199} ({\tiny $\blacksquare$}),
and \cite{PRD76p072012} ({\color{green}$\bullet$}),
with the line shape resulting from Eq.~(\ref{bwpbg}) \lijntje{1}{0}{0},
using the BW parameters of Table~\ref{piekjes}.
On the horizontal axis we mark out the measured
relative kaon momentum in the $K^{+}K^{-}$ CM system,
and on the vertical axis the corresponding cross section.
(a): Non-Resonant contribution \lijntje{0.6}{0.6}{0.6} and
sum of non-resonant and  $\rho$ meson contributions.
(b): Including also the $\phi (1020)$ contribution.
(c): The invariant-mass region of $\phi (1020)$.
(d): Contributions of further resonances.
}
\label{phi}
\end{figure}

Cross sections measured for the reaction $e^{+}e^{-}\to K^{+}K^{-}$
date back to the 1970s.
Here, we shall compare Eq.~(\ref{bwpbg}) to data
from the ORSAY-DCI-DM1 experiment
for invariant masses ranging from 1.4 GeV to 2.06 GeV,
published in 1980 \cite{PLB99p257},
data measured by the NOVOSIBIRSK-OLYA experiment
in the energy range 1.017--1.4 GeV,
published in 1981 \cite{PLB107p297},
more recent data from the CMD-2 Collaboration
determining the $\phi (1020)$ mass and width
\cite{PLB364p199}, published in 1995,
and very recently published (2007) data
obtained with the SND detector in the VEPP-2M experiment,
for invariant masses ranging from 1.04 GeV to 1.38 GeV
\cite{PRD76p072012}.
The results of Eq.~(\ref{bwpbg}) for $r_{0}=0.56$ fm
are shown in Fig.~\ref{phi}.
The BW parameters of the four structures of sufficient significance
in our analysis of the data are collected in Table~\ref{piekjes}.

Although we expected the $KK^{\ast}$ channel
to be the main competitor for the production
of charged-kaon pairs,
we do not even observe any relevant decrease
in the $K^{+}K^{-}$ cross section at the $KK^{\ast}$ threshold.
On the contrary, the non-resonant contribution
to the $K^{+}K^{-}$ line shape is maximum at the $KK^{\ast}$ threshold.
Indeed, from a detailed study performed by the BaBar Collaboration
in Ref.~\cite{PRD77p092002}
we learn that, while the individual isoscalar and isovector
cross sections for $e^{+}e^{-}\to K^{\ast}K$ are comparable
in magnitude to those for $e^{+}e^{-}\to K^{+}K^{-}$,
the total cross section for $e^{+}e^{-}\to K^{+}K^{-}\pi^{0}$
is very small, hinting at destructive interference
between isoscalar and isovector contributions.
Moreover, several other channels compete with the production
of $K^{+}K^{-}$, like
$\eta\omega$, $\eta\phi$, $\eta '\omega$, $\eta '\phi$,
$\pi\rho$ in the isoscalar sector,
and $\eta\rho$, $\eta '\rho$, $\rho\rho$
in the isovector sector.
Several of these channels are in reality smeared out, as they involve
the broad $\rho$ resonance.
Consequently, their effect is spread out over a relatively large
interval of invariant $K^{+}K^{-}$ masses.
The strongest and sharpest channel is probably $\eta '\phi$,
with threshold at almost 2.0 GeV.
Hence, it is not a real surprise that our simple rule of thumb
for the quantity defined in Eq.~(\ref{patBBstar}) does not work in
this case.
For the non-resonant contribution to the theoretical cross section
shown in Fig.~\ref{phi}, we have chosen $r_{0}=0.56$ fm, which is equal
to the kaon charge radius.

A further consequence of the apparent absence of channels
competing with $K^{+}K^{-}$ production
is that it allows us to analyse the experimental cross section
well above threshold by using Eq.~(\ref{bwpbg}),
thereby assuming that no relevant inelasticity occurs.

In Fig.~\ref{phi}a we show the two dominant ``background'' contributions,
viz.\ the non-resonant term of Eq.~(\ref{bwpbg})
and the contribution of the $\rho$ resonance.
The $\phi$(1020) resonance is included in Fig.~\ref{phi}b,
using the PDG resonance parameters \cite{PLB667p1}.
In Fig.~\ref{phi}c only the $\phi$(1020) signal is shown.
The theoretical cross section is adjusted to the data
in Fig.~\ref{phi}d by introducing three resonant structures, namely
(also see Table~\ref{piekjes})
$X(1250)$ ($M=1.25$~GeV, $\Gamma=60$~MeV),
$X(1590)$ ($M=1.59$~GeV, $\Gamma=60$~MeV),
and $X(1830)$ ($M=1.83$~GeV, $\Gamma=80$~MeV).
Additionally, we display in Fig.~\ref{phi}d the contribution of the
non-resonant term of Eq.~(\ref{bwpbg})

\begin{table}[htbp]
\begin{center}
\begin{tabular}{||c||cc||}
\hline\hline & & \\ [-7pt]
piek & Mass (GeV) & Width (MeV)\\
& & \\ [-7pt]
\hline & & \\ [-7pt]
$\phi (1020)$ & 1.01946 & 4.26 \\ [10pt]
$X(1250)$ & 1.25 & 60 \\ [10pt]
$X(1590)$ & 1.59 & 60 \\ [10pt]
$X(1800)$ & 1.83 & 80 \\ [10pt]
\hline\hline
\end{tabular}
\end{center}
\caption[]{\small BW parameters used for
a fit to the data shown in Fig.~\ref{phi}.
The $\phi (1020)$ parameters are taken from Ref.~\cite{PLB667p1}.
}
\label{piekjes}
\end{table}

One observes from Fig.~\ref{phi}
that data for the reaction $e^{+}e^{-}\to K^{+}K^{-}$
roughly follow the tail
of the non-resonant term of Eq.~(\ref{bwpbg})
for $K^{+}K^{-}$ momenta larger than about 0.6 GeV,
which corresponds to a total invariant mass of 1.56 GeV.
Below this mass, the $\phi (1020)$ and $\rho$ resonances dominate
the cross section.
The $\phi (1020)$ line shape is well reproduced
by our theoretical curve, which uses the PDG \cite{PLB667p1}
values for its mass and width.

The recent data of Ref.~\cite{PRD76p072012} are in conflict
with the almost three decades older data of Ref.~\cite{PLB107p297}.
Our theoretical curve agrees better with the former data
for $p\left( K^{+}K^{-}\right)\approx 0.19$ GeV,
which corresponds to an invariant mass of 1.07 GeV.
In particular, three subsequent data points of Ref.~\cite{PLB107p297}
do not fit in the amplitude of Eq.~(\ref{bwpbg}).
However, in the invariant-mass region of 1.17--1.35 GeV
($p\left( K^{+}K^{-}\right)\approx 0.30$--0.45 GeV),
we notice that the data of Ref.~\cite{PRD76p072012}
are substantially larger than those of Ref.~\cite{PLB107p297}
and, moreover, that some of the structure has flattened out.
We should have no difficulty in fitting the data of
Ref.~\cite{PRD76p072012}, as it would only change the parameters of
the $X(1250)$ resonance.
Here, we have opted for reproducing the data of Ref.~\cite{PLB107p297}
in the invariant-mass region of 1.17--1.35 GeV.

\section{Conclusions}
\label{finalities}

We have shown that our recently developed amplitude for
hadron-pair production in electron-positron annihilation
describes quite well the corresponding experimental data for open
strange\-ness, open charm, and open beauty.
Moreover, the formalism allows for a well-defined separation
in resonant and non-resonant contributions.
By choosing the value 2.5 for the quantity
that defines the effectively available window for production,
given in Eq.~(\ref{patBBstar}), we tune the only free parameter $r_{0}$
controlling the non-resonant contributions to data
for $D\bar{D}$, $B\bar{B}$, $\Lambda_{c}^{+}\Lambda_{c}^{-}$,
and $K^{+}K^{-}$. As a result, we obtain for $r_0$ a value close to
1 fm for the former three process, whereas for the latter one the kaon
charge radius turns out to work well.
Finally, it follows from our approach that the size of the window
rather accurately defines the total amount of non-resonant
hadron-pair production.

In the $D\bar{D}$ case, we find,
besides a substantial non-resonant contribution,
a narrow $\psi (3770)$ resonance,
with a central mass of 3.781 GeV and a width of 17 MeV,
both well within the error bars
suggested by the BES Collaboration.
As for the bottomonium sector, the $\Upsilon (10580)$ enhancement
in $B\bar{B}$ appears to contain exclusively non-resonant contributions,
so that it should not be associated with a $b\bar{b}$ state.
A wave function with a large $B\bar{B}$ component is likely to be
a more appropriate description of this enhancement.

For the cross section of $\Lambda_{c}^{+}\Lambda_{c}^{-}$ production,
one needs to assume,
besides the non-resonant enhancement near threshold,
two new $c\bar{c}$ resonances, which are naturally interpreted as the
$\psi (5S)$ and the $\psi (4D)$.
We obtain 4.784 GeV and 4.871 GeV
for their central resonance masses,
and 55 MeV and 60 MeV for their widths, respectively.
The non-resonant contribution to $K^{+}K^{-}$ production
is rather modest.
Apart from a very accurate description of the $\phi (1020)$,
we find three further enhancements that can be fitted as BWs,
namely the $X(1250)$, $X(1590)$, and $X(1830)$,
with their central masses
at 1.25 GeV, 1.59 GeV, and 1.83 GeV, respectively,
and widths of 60--80 MeV.

\section*{Acknowledgments}

We are grateful for the precise measurements
and data analyses of the various experimental Collaborations
that made the present analysis possible.
We also thank Dr.\ Xiang Liu  of Lanzhou University for drawing our attention
to the recent results of the BES Collaboration
concerning the $\psi (3770)$ resonance,
and Dr.\ K.~P.~Khemchandani for useful discussions.
This work was supported in part by
the \emph{Funda\c{c}\~{a}o para a Ci\^{e}ncia e a Tecnologia}
\/of the \emph{Minist\'{e}rio da Ci\^{e}ncia, Tecnologia e Ensino Superior}
\/of Portugal, under contract CERN/\-FP/\-83502/\-2008.

\newcommand{\pubprt}[4]{{#1 {\bf #2}, #3 (#4)}}
\newcommand{\ertbid}[4]{[Erratum-ibid.~{#1 {\bf #2}, #3 (#4)}]}
\def\AIPCP{AIP Conf.\ Proc.}
\def\AP{Ann.\ Phys.}
\def\EPL{Europhys.\ Lett.}
\def\JPG{J.\ Phys.\ G}
\def\PLB{Phys.\ Lett.\ B}
\def\PRC{Phys.\ Rev.\ C}
\def\PRD{Phys.\ Rev.\ D}
\def\PRL{Phys.\ Rev.\ Lett.}
\def\ZPC{Z.\ Phys.\ C}

\end{document}